\documentclass[apj,twocolumn]{emulatearxiv}
\usepackage{amsmath}

\begin{document}

\title{Pulsational Pair-instability Supernovae. II. Neutrino Signals from Pulsations
and their Detection by Terrestrial Neutrino Detectors}

\author{
Shing-Chi Leung\thanks{Email address: shingchi.leung@ipmu.jp}}

\affiliation{Kavli Institute for the Physics and 
Mathematics of the Universe (WPI),The University 
of Tokyo Institutes for Advanced Study, The 
University of Tokyo, Kashiwa, Chiba 277-8583, Japan}

\affiliation{TAPIR, Walter Burke Institute for Theoretical Physics, 
Mailcode 350-17, Caltech, Pasadena, CA 91125, USA}

\author{Sergei Blinnikov\thanks{Email address: sblinnikov@gmail.com}}

\affiliation{Kavli Institute for the Physics and 
Mathematics of the Universe (WPI),The University 
of Tokyo Institutes for Advanced Study, The 
University of Tokyo, Kashiwa, Chiba 277-8583, Japan}

\affiliation{NRC "Kurchatov Institute" -- ITEP, B.Cheremushkinkaya 25, 117218 Moscow, Russia}

\affiliation{Dukhov Automatics Research Institute (VNIIA), Suschevskaya 22, 127055 Moscow, Russia}

\author{Koji Ishidoshiro\thanks{Email address:koji@awa.tohoku.ac.jp}}

\affiliation{Research Center for Neutrino Science, 
Tohoku University, Sendai 980-8578, Japan}

\author{Alexandre Kozlov\thanks{Email address:kozlov@awa.tohoku.ac.jp}}

\affiliation{Kavli Institute for the Physics and 
Mathematics of the Universe (WPI),The University 
of Tokyo Institutes for Advanced Study, The 
University of Tokyo, Kashiwa, Chiba 277-8583, Japan}

\author{Ken'ichi Nomoto\thanks{Email address: nomoto@astron.s.u-tokyo.ac.jp}}

\affiliation{Kavli Institute for the Physics and 
Mathematics of the Universe (WPI),The University 
of Tokyo Institutes for Advanced Study, The 
University of Tokyo, Kashiwa, Chiba 277-8583, Japan}

\date{\today}

\begin{abstract}

A Pulsational Pair-instability supernova (PPISN) evolves from a massive star
with a mass $\sim 80$ -- 140 $M_{\odot}$ which develops the electron-positron
pair-instability after the hydrostatic He-burning in the core has finished. 
In [Leung et al., ApJ \textbf{887}, 72 (2019)] (Paper I) we examined the evolutionary
tracks and the pulsational mass loss history of this class of stars. 
In this paper, we analyze the thermodynamical history to explore the neutrino observables of PPISNe.
We compute the neutrino light curves and spectra during pulsation.
We study the detailed neutrino emission profiles of these stars.
Then, we estimate the expected neutrino detection count for different 
terrestrial neutrino detectors including, e.g., KamLAND and Super-Kamiokande.
Finally, we compare the neutrino pattern of PPISN with other types of supernovae
based on a canonical 10 kt detector. 
The predicted neutrino signals can provide the early warning
for the telescopes to trace for the early time optical signals. 
Implications of neutrino physics on the expected detection are discussed.

\end{abstract}




\section{Introduction}

\subsection{Pulsational Pair-Instability Supernova}

Pulsational pair-instability supernova (PPISN)
is the explosion of a massive star by the instabilities 
during its pulsation. This occurs in a
star with a mass from $\sim 80$ to $\sim 140$ $M_{\odot}$, where the exact mass
is metallicity dependent. After He-burning,
the massive C+O core experiences pair-creation instabilities \citep{Barkat1967}, 
where energetic photons which support the star 
are forming electron-positron pairs catastrophically
during its contraction. 
Such a core can form when the metallicity is 
sub-solar ($\sim 0.8 ~Z_{\odot}$), where the massive star can develop
a He-core above 40 $M_{\odot}$. The stellar wind mass loss 
is suppressed during the main-sequence phase \citep{Hirschi2017,Limongi2017,Leung2018}.
The conversion of photons drastically lowers the 
radiation pressure, making the adiabatic index $< 4/3$.
This makes the star enter an over-compressed
state. Explosive O-burning is triggered which makes the star
rebounce and pulsate. Depending on the pulsation
strength, which increases with the stellar mass, a PPISN may eject
a significant fraction of mass. After that the star 
expands and relaxes. The star gradually contracts
by losing energy through radiation and neutrinos \citep{Woosley2017},
after that the star resumes its contraction. Depending on 
the amount of unburnt O left behind by the previous explosive
O-burning and its replenishment from the outer zone by
convective mixing, the star can 
carry out the above process repeatedly until the core runs
out of O. At that point, the star collapses as a
core collapse supernova (CCSN). The combination of thermonuclear
runaway and core-collapse in one single star makes this class
of stars interesting. 
We refer readers to \cite{Heger2002,Ohkubo2009,Yoshida2016b,Woosley2017,
Woosley2018,Marchant2018,Leung2019}
for some recent calculations of the PPISN pulsations and progenitor modeling of PPISNe. 

PPISN is less studied than other types of supernovae
due to its numerical complication. 
It contains dynamical phases and quiescent phases.
The dynamical phase occurs during
pulsations where the dynamical timescale becomes shorter 
than the nuclear reaction timescale. 
The quiescent phase occurs 
between pulses, where the 
Kelvin-Helmholtz timescale determines the contraction time.
It is 
difficult to follow its overall evolution with a 
single code. Multiple codes are used (see e.g. \cite{Yoshida2016b})
or an excerpt of the pulsation is followed (see e.g. \cite{Chen2014}).
Recent development of the stellar evolution code MESA 
(Modules for the Experiments in Stellar Astrophysics) 
\citep{Paxton2011,Paxton2013,Paxton2015,Paxton2017}
allows flexible changes between the hydrostatic approximation for the 
quiescent phase and the implicit hydrodynamics calculations for the 
dynamical phase. 

Despite the difficulty, PPISNe are important because they are one of the robust
mechanisms for producing super-luminous supernovae. The massive mass loss
during pulsation creates a rich circumstellar medium (CSM).
They are also one of the channels for forming massive black holes ($\sim 30 - 50 ~M_{\odot}$)
where the merger events of such black holes can generate gravitational wave signals detected
by for example advanced LIGO and VIRGO \citep{Belczynski2017}.
During the final explosion when the Fe-core collapses,
the ejecta interacts with the CSM and creates shock breakout.
Such process can produce a very bright event to explain some
super-luminous supernovae including e.g. PTD12dam \citep{Sorokina2016,Tolstov2017},
Eta Carinae \citep{Woosley2017} and iPTF14hls \citep{Woosley2018}.

\subsection{Neutrino as Another Messenger of Supernovae}

In this and the coming decades, the increasing size of neutrino
detectors has made probing neutrinos from the astrophysical 
sources possible, for example, the upgrade of
the Super-Kamiokande to the Hyper-Kamiokande increases the detection mass
from 32.5 kton \citep{Simpson2019} to 220 kton, which is expected 
to be realized in the later half of 2020s. 
In terms of the energy range the large neutrino detector ICECUBE 
can detect neutrinos with an energy up to PeV. This enables the detection
of neutrino sources beyond the Sun and CCSNe to objects such as blazars
and supernova remnants in compact stellar clusters \citep{Bykov2015}.
The low neutrino interaction cross-section with matters 
allows supernova neutrinos likely to reach the Earth before photons. 
The arrival of neutrinos, if detected, can serve as an
early warning signal, used by the SuperNova Early Warning System 
(SNEWS) \citep{Antonioli2004}. 
Gravitational wave signals can also serve as a similar early warning for 
merging compact stars. The gravitational wave signal is significant
in a binary system but it is much weaker in the single star scenario. 
On the other hand, the neutrino emission can be significant
in both a single star event, during its thermonuclear explosion or its core-collapse,
and in binary star interactions such as a binary neutron star merger event.

Large terrestrial detectors are built or proposed
including 
1. liquid scintillator detectors (e.g.
the Kamioka Liquid-scintillator Antineutrino Detector
(KamLAND) in Japan \citep{Suzuki1999,Asakura2016}, SNO+ in Sudbury, Canada \citep{Andringa2016}, 
Boron solar neutrino experiment (Borexino) in Gran Sasso, Italy 
\citep{Agostini2015, Bellini2014}, 
The Jiangmen Underground Neutrino Observatory (JUNO) in China, \citep{An2016}, 
RENO-50 in Korea \citep{Seo2015},
and Low Energy Neutrino Astronomy (LENA) in Europe \citep{Wurm2012});
2. Water Cherenkov detector in Super-Kamiokande and Hyper-Kamiokande in Japan
\citep{Watanabe2009, Abe2011a,Abe2011b} and IceCube in South Pole \citep{Abbasi2011}, 
3. Gadolinium-loaded water Cherenkov detectors Super-Kamiokande
and Hyper-Kamiokande \citep{Beacom2004}, and 
4. liquid argon detector (the Deep Underground Neutrino Experiment - DUNE)
in the USA \citep{Acciarri2016}. 
In Table \ref{table:detectors} we provide more specific details on 
the characteristics of these detectors. Detectors designed to detect
both electron and anti-electron neutrinos are included for comprehensiveness.
These detectors provide a wide range of exposure cross sections
for multiple types of neutrinos and reaction channels. 
The diverse locations of neutrino detectors allow 
measuring the supernova position by the time-delay between
neutrino detection among various detectors.
Detection methods such as the 
triangulation method \citep{Brdar2018} are proposed to
identify the neutrino source to a sub-degree accuracy.
However, it also requires absolute time synchronization between 
detectors and knowledge of the arrival time of the neutrino pulses. 
The limited number of events in each detector may cap the accuracy in 
determining the arrival time of the neutrino signals.

So far, neutrino signals from thermonuclear related
supernovae have been largely studied, 
including Type Ia supernovae
(see e.g. \cite{Kunugise2007,Odrzywolek2011,Wright2016,Wright2017a}) 
and PISN \citep{Wright2017b}. CCSN
is also a natural source of astrophysical neutrino but 
electron captures, neutron star cooling and
its accretion are the major production mechanisms.
The neutrino signal
contains information about the core \citep{Suwa2019}, 
which may complement to its optical observable where
the mass ejection occurs on the surface.
As remarked above, PPISN could be an important 
source of neutrinos due to its lower mass compared to 
PISN while having a significant thermonuclear
burning during pulsation. Its multiple pulses also 
offer more chances to produce neutrinos compared to the 
single explosive event in the other two types of supernovae. 

\begin{table*}
\begin{center}
\caption{The characteristics of some recent neutrino detectors.}
\begin{tabular}{|c|c|c|c|c|c|c|}
\hline
detector & location & mass (kt) & detection type & medium & main neutrino detected & others \\ \hline
KAMLAND & Japan & 1 & liquid scintillator & organic liquid & $\bar{\nu}_e$ & \\
SNO+ & Canada & 0.78 & liquid scintillator & organic liquid & $\nu_e$ & \\
Borexino & Italy & 0.278  & liquid scintillator & organic liquid & $\nu_e$ & \\ 
JUNO & China & 20 & liquid scintillator & organic liquid & $\bar{\nu}_e$ & \\
RENO-50 & Korea & 18 & liquid scintillator & organic liquid & $\bar{\nu}_e$ & \\ \hline
Super-Kamiokande & Japan & 32.5 & Water Cherenkov detector & H$_2$O & $\bar{\nu}_e$ & With Gd \\ 
Hyper-Kamiokande & Japan & 220 & Water Cherenkov detector & H$_2$O & $\bar{\nu}_e$ & With Gd \\ \hline
DUNE & USA & 40 & liquid argon detector & Liquid Ar & $\nu_e$ & \\  \hline
\end{tabular}
\label{table:detectors}
\end{center}
\end{table*}

\subsection{Motivation}

To our knowledge
there is not yet any systematic study about neutrino
signals from PPISNe. In this work, we explore the neutrino 
signature including the neutrino luminosity and spectra
based on our PPISN evolutionary models
computed by MESA. We present our study about the typical
features of neutrino signals emitted during pulsations 
in this class of supernovae. 

In Section \ref{sec:methods}
we describe the code we used for 
preparing for the stellar models and then we 
describe the numerical scheme for extracting 
the neutrino light curves and spectra. 
In Section \ref{sec:neutrino} 
we present in details the neutrino emission
profiles and thermodynamical history of these models. 
In Section \ref{sec:discussion} we
predict the expected neutrino 
detection rates by the existing and proposed
neutrino detectors. Then we compare the neutrino 
pattern with other types
of supernovae.
At last we give our conclusions.
In Appendix \ref{sec:analytic} we compare the 
use of the analytic approximation to the numerical
scheme we have used for calculating the neutrino luminosity. 

\section{Methods}
\label{sec:methods}

For the hydrodynamics model, we refer the interested readers
to \cite{Leung2019} (Paper I) for the detailed implementation. 
We have used the stellar evolution code MESA version 8118
(Modules for the Experiments in Stellar Astrophysics)
\citep{Paxton2011, Paxton2013, Paxton2015, Paxton2017}
for computing the PPISN models from the main-sequence
phase until the onset of Fe-core collapse. The implicit hydrodynamical
scheme is used for following the pulsation of the star until
the mass ejection is finished. 

To reconstruct the neutrino emission history, we use the neutrino
energy loss subroutine provided in MESA. It accounts for several
major neutrino emission channels including pair-, photo-,
plasma and bremmstrahlung neutrinos, where the analytic formulas
are given in \cite{Itoh1989} \footnote{Open-source subroutines
are available in the website http://cococubed.asu.edu/}.
To calculate the neutrino spectra, we use the formulas
given in \cite{Misiaszek2006,Odrzywolek2007}, which 
contain the pair-annihilation and plasma neutrinos. 
We refer the readers to the original articles 
for the derivation of these formulas.

The number emission of the pair-neutrinos $\phi_{{\rm pair}} ( \epsilon )$  is given
by the approximation
\begin{equation}
\phi_{\rm pair} (\epsilon) = \frac{A}{k_B T} \left( \frac{\epsilon}{k_B T} \right)^{\gamma} \exp (- a \epsilon / k_B T).
\label{eq:pairspec}
\end{equation}
Variables $\alpha$, $a$ and $A$ are fitting
parameters where $\alpha = 3.180657028$, $a = 1.018192299$ and 
$A = 0.1425776426$. 
Notice that the fitting here assumes the matter being 
relativistic and non-degenerate, i.e. $k T > 2 m_e$ 
and $k T > \mu_e$. In the pre-supernova scenario, 
such conditions may not be always satisfied. However, 
we argue that such approximation will have small 
effects because its number emissivity scales
directly with the total emissivity,
which is dominated by the pair-annihilation rate. 
The emission spectrum is to a good approximation a
thermal spectrum. 

To calculate the plasmon neutrino spectrum and 
emissivity, we follow the prescription from
\cite{Odrzywolek2007}, where
\begin{equation}
\phi_{{\rm plasmon}} = A k_B T m_t^6 \exp(-\epsilon / k_B T),
\end{equation}
and
\begin{equation}
A = \frac{G_F^2 C_V^2}{8 \pi^4 \alpha}.
\end{equation}
In cgs units, $A = 2.115 \times 10^{30}$ MeV$^{-8}$ cm$^{-3}$ s$^{-1}$. 
Notice that one needs to take the corresponding $C_V$ 
for electron-neutrinos and muon-/tau- neutrinos respectively
for calculating pair-neutrinos. 

The asymptotic transverse plasmon mass $m_t$ is given by 
\begin{equation}
m_t = \frac{4 \alpha}{\pi} \int_0^{\infty} \frac{p^2}{E} (f_1 + f_2) dp,
\end{equation}
where 
\begin{equation}
f_i = \frac{1}{1 + \exp[(E + (-1)^i \mu)/ k_B T]},
\end{equation}
which represents the Fermi-Dirac distributions of
electrons $(i = 0)$ and positrons $(i = 1)$. In general, 
plasmon-neutrino is a less significant neutrino 
source compared to the pair-neutrino in the 
thermodynamics range we are interested. 

\section{Neutrino Signals}
\label{sec:neutrino}

\subsection{Review of Hydrodynamics Results}
\label{sec:hydro}

\begin{table*}
\begin{center}

\caption{The stellar evolutionary models prepared by the MESA code.
$M_{{\rm ini}}$ and $M_{{\rm fin}}$ are the initial and final 
masses of the star. $M_{{\rm He}}$ and $M_{{\rm CO}}$
are the integrated helium and carbon-oxygen 
masses of the whole star before the
dynamical phase starts. No hydrogen mass is given
because we start the star as a bare He core.
"Weak Pulse" and "Strong Pulse" refer
to the numbers of the corresponding pulses in the evolutionary 
history. All masses are in units of 
solar mass.}
\label{table:He_MS_result}
\begin{tabular}{|c|c|c|c|c|c|c|c|c|c|}
\hline

Model & $M_{{\rm ini}}$ & $M_{{\rm fin}}$ & $M_{{\rm He}}$ & $M_{{\rm C}}$ & $M_{{\rm O}}$ & Weak Pulse & Strong Pulse &  Ejected mass  \\ \hline
He40A & 40 & 37.78 & 6.79 & 3.13 & 27.5 & 5 & 1 & 2.22 \\
He45A & 45 & 39.26 & 7.38 & 4.03 & 31.3 & 3 & 1 & 5.74 \\
He50A & 50 & 47.39 & 7.82 & 4.16 & 35.2 & 1 & 1 & 2.61 \\
He55A & 55 & 48.22 & 8.27 & 4.30 & 39.0 & 1 & 1 & 6.78 \\
He60A & 60 & 51.48 & 8.69 & 4.43 & 42.9 & 0 & 2 & 8.52 \\
He62A & 62 & 49.15 & 8.77 & 4.59 & 44.6 & 0 & 2 & 12.85 \\
He64A & 64 & 0 & 8.96 & 4.63 & 46.1 & 0 & 1 & 64.00 \\ \hline

\end{tabular}
\end{center}
\end{table*}

First, we review the hydrodynamics properties of the 
PPISNe presented in Paper I. In that work, 
we have followed the evolution of the He cores from
40 $M_{\odot}$ to 64 $M_{\odot}$ from the main-sequence
phase until the onset of core-collapse using the MESA
code. The pure He core assumes no metal at the beginning, 
thus resembling with zero metallicity models. However, 
the metallicity does not affect the pulsation strength 
of a given He core mass, because it depends on the 
electron-positron pair-creation instabilities
and the energy production of the explosion O-burning.
These stars develop and form PPISNe after the massive He cores
with masses $> 40 ~M_{\odot}$ have formed. 
However, we remark that whether the star can form 
the He core massive enough for the PPISN event to occur
depends on its mass loss rate, which is dependent upon 
the stellar metallicity. When these stars are in a binary
system, interaction with its companion star can affect
the final He core mass before the onset of the pair-instability \citep{Marchant2018}. 
As reported in Paper I, the final He core mass 
can be as low as $30 ~M_{\odot}$ at solar metallicity, 
up to $45 ~M_{\odot}$ at one-tenth of solar metallicity.

The quasi-hydrostatic approximation is used
for most parts of the simulations. Implicit hydrodynamics 
formalism is used while following the pulsation and mass-ejecting
phases. 

In all models, we classified two classes of pulsations:
weak pulses and strong pulses. A weak pulse is
the expansion of the core without any mass
loss, while a strong pulse is that with
mass loss. A weak pulse occurs often in a low mass 
He core (below 50 $M_{\odot}$). Above 50 $M_{\odot}$,
the first explosive O-burning is always strong enough to eject part
of the surface, or even matter in the CO layer. For a low
mass He core with $M_{{\rm He}} < 55 ~M_{\odot}$, 
the pulsation can only eject about 1 -- 2
$M_{\odot}$ overall. For more massive He cores, 
especially those close to the PISN limit ($M_{{\rm He}} \sim 64 ~M_{\odot}$),
a mass ejection above 10 $M_{\odot}$ is possible.
Accompanying with the pulsations, the stellar luminosity can 
be 3 -- 4 orders of magnitude higher than that
during the quiescent phase.  

In general, the number of weak pulses decreases when 
the He core mass increases. Conversely, there are more strong
pulses when the He core becomes more massive. It is because,
when the He core mass is closer to the pair-instability regime 
(i.e. 64 $M_{\odot}$), the softening of the C+O core after
the hydrostatic He burning is more significant. The level
of compression until bounce, the amount of C+O matter burnt 
in the process and the released energy are higher. Thus the
strength of the pulse increases, which is more likely to eject
more mass. 
We also refer Paper I for the detailed physics
of the pulsation history.

In all models, we treat $t = 0$ to be 
the moment when the first switch to implicit hydrodynamics starts.
The switch to hydrodynamics is determined by the current timestep that 
the timestep is comparable with the Courant 
timestep. That means, when the onset of the pair-creation instability starts,
the dynamical time gradually decreases as the density of the star
increases. It becomes comparable or even shorter than the nuclear
reaction timescale during the pulsation phase.

Different from stars of higher or lower masses, the
core of a PPISN can reach above $10^9$ K and then fall below that more than
once as long as it pulsates, with its central density ranging from
$\sim 10^5$ to $10^7$ g cm$^{-3}$. The whole process
can last for $\sim$ 1 hour, and the hot stellar core
emits an abundant amount of thermal neutrinos.

\begin{figure}
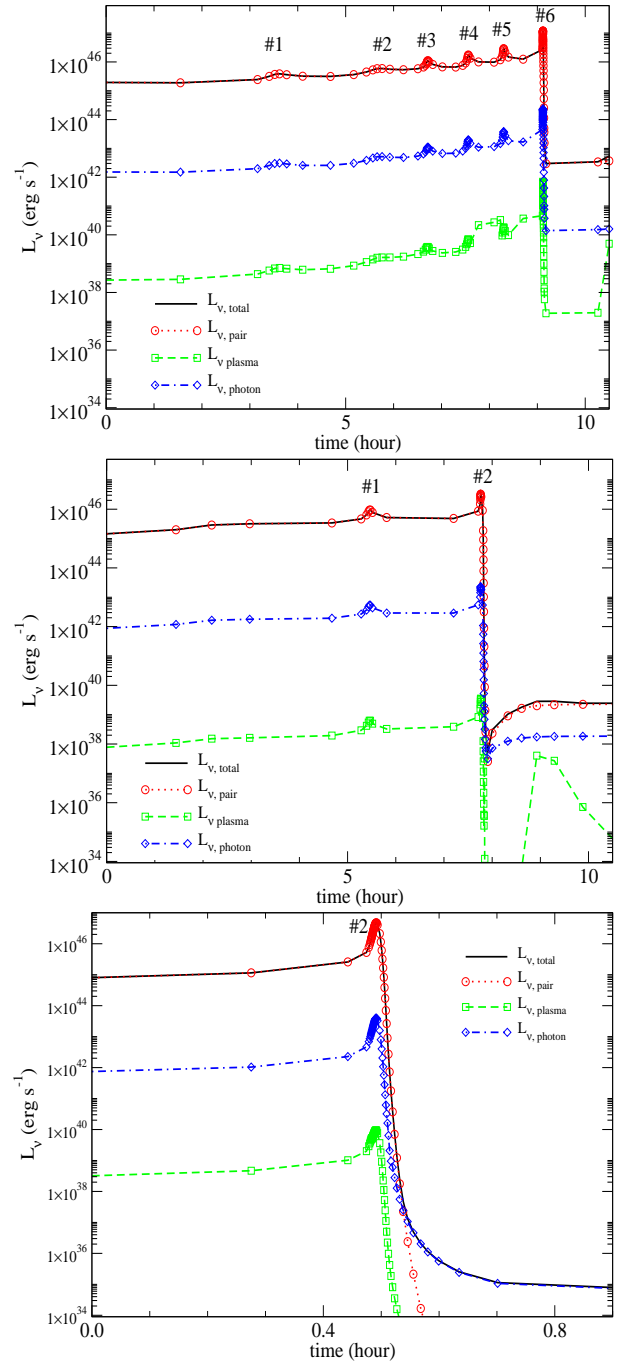

\centering
\includegraphics*[width=8cm,height=6cm]{fig1a.eps}
\includegraphics*[width=8cm,height=6cm]{fig1b.eps}
\includegraphics*[width=8cm,height=6cm]{fig1c.eps}
\caption{
(top panel) The neutrino luminosity and its components
against time for Model He40A, including 
the pair-, plasma and recombination neutrino. All the pulses
before collapse are included. 
(middle panel) Same as the top panel, but for Model He50A
for all the pulses.
(bottom panel) Same as the top panel, but for Model He62A
for the final pulse.
In all three panels, zero time is defined by the 
start of the hydrodynamics, i.e. the beginning where
the star becomes dynamical as it enters the pair creation 
instabilities.
}
\label{fig:nulumin_plot}
\end{figure}

\subsection{Neutrino Luminosity}
\label{sec:nu_signal}

In this section, we post-process the thermodynamics data
from the simulations using MESA by the 
analysis described in Section \ref{sec:methods}.
This means, based on the hydrodynamical results 
reported in Paper I for the density, temperature and composition
profiles of the He cores at different time slices, we reconstruct
the total neutrino emission rates, neutrino luminosity in each
channel and the time-dependent spectra.
We analyze three distinctive models, He40A, He50A and
He62A. We study their neutrino emissitivities, average
neutrino energies and cumulative neutrino emission. 
Specific moments of the neutrino emission profiles
are examined to understand how the star produces
neutrinos. 

The three models represent PPISNe with 
mild, moderate and strong mass losses, which stand
for different levels of mass ejections. 
We remark that Model He40A is interesting because 
it demonstrates
consecutive weak pulses before its last strong pulse.
Such weak pulses largely delay the contraction, which allows
the core to have a higher central density,
which strongly enhances neutrino emission. 
Model He50A demonstrates the standalone strong pulse
with a moderate mass ejection. Model He62A demonstrates
the standalone strong pulse with a significant mass 
ejection near the PPISN-PISN transition. 

In Figures \ref{fig:nulumin_plot} we plot the neutrino luminosities
and their components for Models He40A, He50A and
He62A. The typical neutrino luminosity is about 
$10^{46}$ erg s$^{-1}$ during the peak of the pulse. 
In all pulses, the pair neutrino is the major source
of neutrinos, compared to other channels including the 
photo-neutrino and plasma-neutrino. Photo-neutrinos are
always $\sim$ 2 -- 3 orders of magnitude less than the pair-neutrinos and the 
plasma-neutrinos are another 2 -- 3 orders of magnitude less. This suggests that 
considering only the pair-neutrino gives an accurate estimation of 
the total neutrino luminosity for the pulsations in PPISNe,
which is consistent with that discussed in \cite{Blinnikov1989}. In Appendix A
we present a more detailed comparison between the tabular form and 
the analytic rates. 
During the quiescent phase the neutrino luminosity is negligible 
compared to its peak values, which can be 4 -- 10 orders of 
magnitude higher.

\subsection{Neutrino Spectra during Pulsation}

\begin{figure}
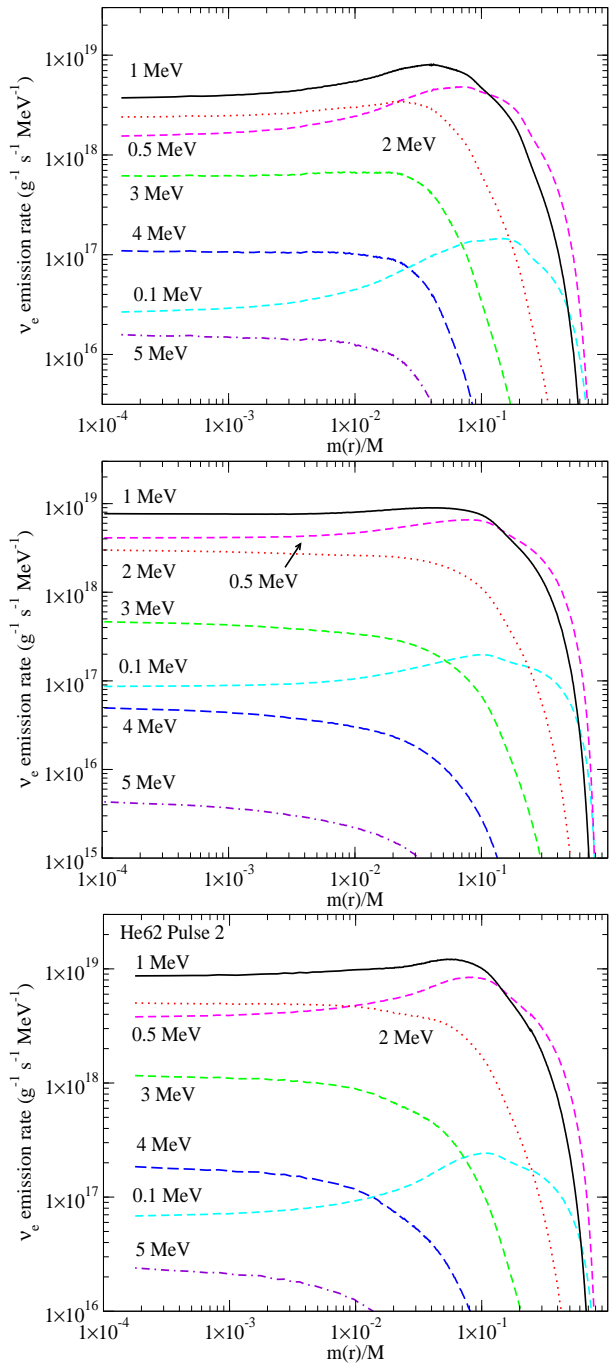

\centering
\includegraphics*[width=8cm,height=6cm]{fig2a.eps}
\includegraphics*[width=8cm,height=6cm]{fig2b.eps}
\includegraphics*[width=8cm,height=6cm]{fig2c.eps}
\caption{
(top panel) The neutrino spectra of Model He40A at the peak 
of the final pulse for neutrinos with energies from 
0.1 -- 5 MeV.
(middle panel) Same as the top panel, but at the peak of 
the final pulse for Model He50A.
(bottom panel) Same as the left panel, but at the peak of 
the final pulse for Model He62A.
}
\label{fig:nuprofile_plot}
\end{figure}

In Figures \ref{fig:nuprofile_plot} we plot the neutrino spectra 
of neutrinos with energies from 0.1 to 5 MeV,
during
the peak of the pulses for the three mentioned models. 
The spectrum is single snapshot obtained by integrating the neutrino
emission in the whole star when the neutrino luminosity
reaches its maximum during a pulse.
The neutrino spectra include contributions of 
both pair-neutrinos and photo-neutrinos. By examining the patterns
of the neutrinos, we can see that the neutrino emission in most
cases remains thermal that the number emission drops when 
the neutrino energy increases. Below 1 MeV, the 
neutrino number drops rapidly. Neutrinos with an energy 0.1 MeV
is almost as low as those with an energy 5 MeV. In general,
the energy threshold of current neutrino detectors is $\sim$ 1 MeV,
the low energy neutrinos are not counted as detection. 
Future generation-3 noble liquid-based neutrino detectors 
using argon, silicon, germanium and xenon
as the scintillator, such as DARWIN \citep{Aalbers2016} and ARGO \citep{Aalseth2018}, can allow 
much lower energy thresholds based on the technique used in
dark matter detection \citep{Raj2019b}.
It increases the chance of capturing supernova neutrinos 
for distinguishing the supernova explosion mechanisms \citep{Raj2019a}.

By comparing the shape of the neutrino spectra,
it shows that the PPISN shares similar neutrino spectra
where low energy neutrinos ($\sim 1$ MeV) dominate
the emission, while higher energy neutrinos ($\sim 5$ MeV)
can be 2 -- 3 orders of magnitude lower. This shows
that during pulsation, the core has only barely reached
the temperature for producing thermal neutrinos. 
Nevertheless, the central temperature can be as hot as $10^{9.5 - 9.7}$ K. 
The neutrino production focuses mostly at 
$q = m(r)/M \approx 0.1$ for all three cases as shown 
by the bumps for 1 -- 2 MeV neutrinos. They are the
places where very active burning takes places.

\subsection{Neutrino Number Evolution during Pulsation}

\begin{figure}
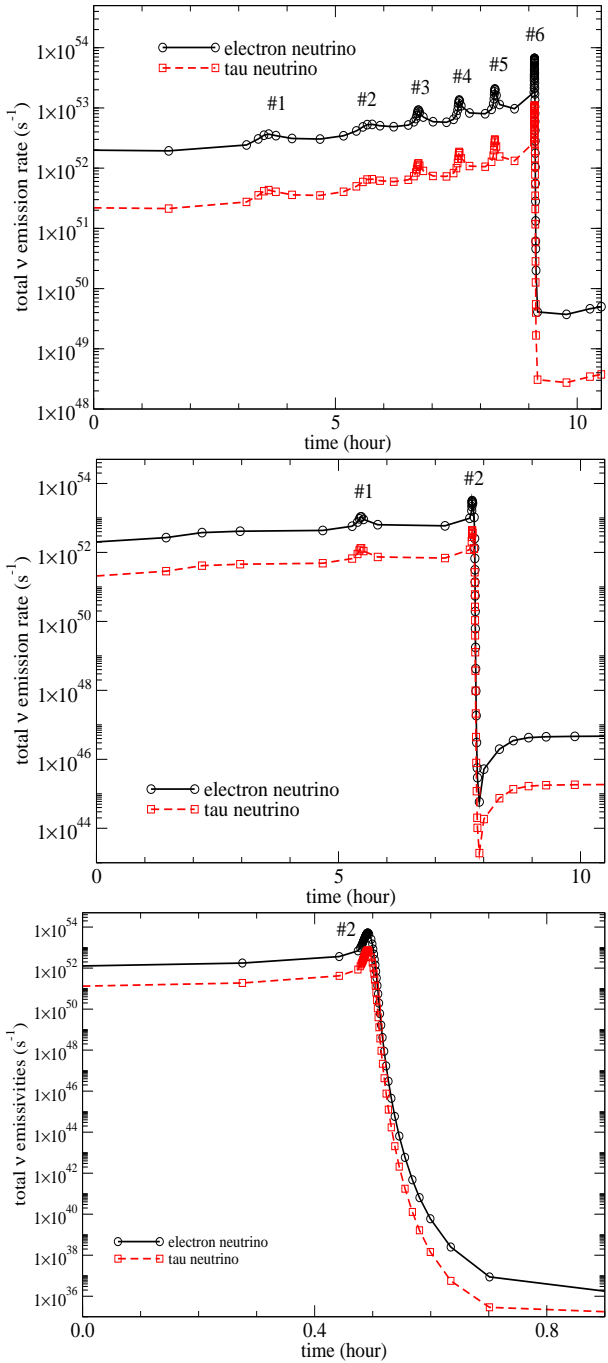

\centering
\includegraphics*[width=8cm,height=6cm]{fig3a.eps}
\includegraphics*[width=8cm,height=6cm]{fig3b.eps}
\includegraphics*[width=8cm,height=6cm]{fig3c.eps}
\caption{
(top panel) The neutrino number emission of Model He40A at the peak 
of all the pulses.
(middle panel) Same as the top panel, but at the peak of 
all the pulses for Model He50A.
(bottom panel) Same as the top panel, but at the peak of 
the final pulse for Model He62A.
The time convention follows Figure \ref{fig:nulumin_plot}.
}
\label{fig:nunum_plot}
\end{figure}

In Figure \ref{fig:nunum_plot} we plot the energy-integrated neutrino
number emission rate for the same set of models at the peaks of
the pulses of Models He40A, He50A and He62A. 
The star emits neutrinos at a rate of $\sim 10^{50}$ s$^{-1}$ when the star contracts
after the core has exhausted its He. Then, it quickly rises
to $\sim 10^{52} - 10^{53}$ s$^{-1}$ when the 
core reaches its maximum compactness. 
Most neutrinos are emitted within $10^{-4}$ year ($\sim$ 1 hour) up
to the temperature peak reached by the core. Then, the neutrino emission quickly 
falls. This means that for most pulses there is only one major outburst of 
neutrinos coming from the core, then the core expands and becomes
too cold for further neutrino emission. 

The duration where most neutrinos are emitted decreases when the progenitor mass increases.
Mode He40A shows an extreme extension. It is because before its final pulse, the weak pulses
do not expand the star or cool down the core. Thus, the neutrino 
emission continues, which provides a longer duration compared
to the other five models. Model He50A shows a sharp peak
of the neutrino flux before expansion. On the other hand Model
He62A shows a smooth but rapid rise and fall in the neutrino 
emission rate. 

\subsection{Neutrino Spectra Evolution}

\begin{figure}
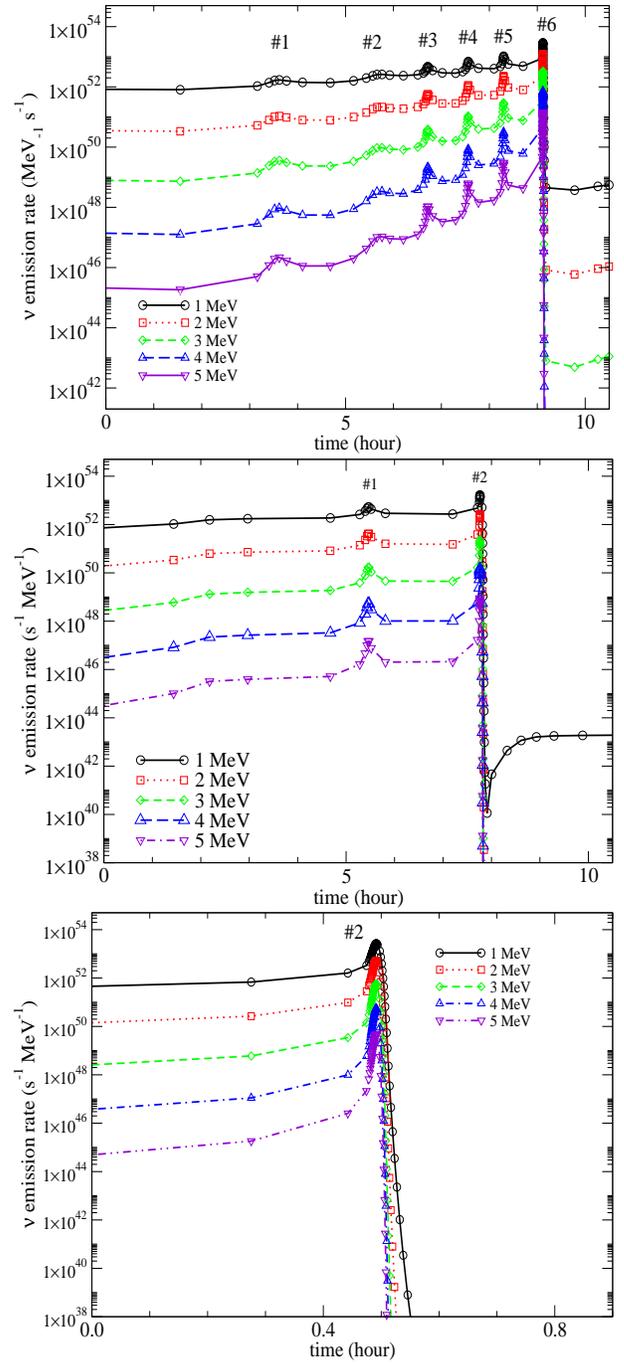

\centering
\includegraphics*[width=8cm,height=6cm]{fig4a.eps}
\includegraphics*[width=8cm,height=6cm]{fig4b.eps}
\includegraphics*[width=8cm,height=6cm]{fig4c.eps}
\caption{
(top panel) The neutrino number emission against time for
neutrino energies from 1 MeV to 5 MeV of Model He40A.
(middle panel) Same as the top panel, but for Model He50A.
(right panel) Same as the top panel, but at the peak of 
the second pulse for Model He62A.
The time convention follows Figure \ref{fig:nulumin_plot}.}
\label{fig:nuspec_plot}
\end{figure}

We examine the evolution of neutrino spectra 
for the three models. In Figure \ref{fig:nuspec_plot}
we plot the neutrino spectra as a function of 
time for the same set of models. 

The typical neutrino number emission for each 
band follows a similar structure. It is because 
they depend on the same scaling relation in 
Eq. \ref{eq:pairspec}. The neutrino emission 
increases during contraction and decreases during
expansion. The typical emission number at the peak 
is $\sim 10^{53}$ s$^{-1}$ MeV$^{-1}$. 
The number emission rate typically drops by one
order of magnitude when the neutrino energy 
increases by 1 MeV. Despite that the shape of the curve
follows each other, showing only thermal contributions.
The contraction in models with a lower He core mass is slower, 
thus the neutrino number emission rate exhibits 
more features. On the other hand, for a more
massive He core, expansion follows immediately
after contraction and the explosive O-burning, 
thus the neutrino signal has only
a one-peak feature.

\subsection{Neutrino Energy Evolution}

\begin{figure}
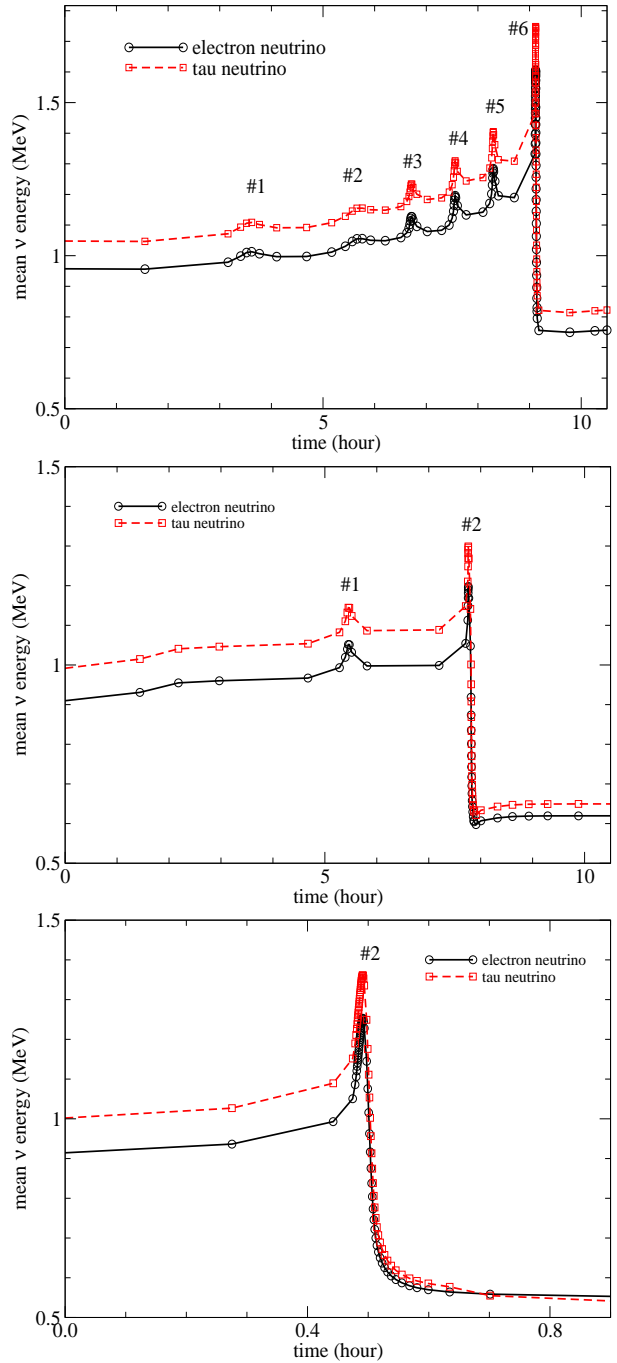

\centering
\includegraphics*[width=8cm,height=6cm]{fig5a.eps}
\includegraphics*[width=8cm,height=6cm]{fig5b.eps}
\includegraphics*[width=8cm,height=6cm]{fig5c.eps}
\caption{
(top panel) Mean neutrino energy against time for
$\nu_e$ and $\nu_{\tau}$ of Model He40A for all the 
pulses.
(middle panel) Same as the top panel, but for Model He50A
for all the pulses.
(bottom panel) Same as the top panel, but at the peak of 
the second pulse for Model He62A during the second strong
pulse.
The time convention follows Figure \ref{fig:nulumin_plot}.}
\label{fig:numean_plot}
\end{figure}

At last, we examine the mean energy of both $\nu_e$ and 
$\nu_{\tau}$ in our models. The mean energy is obtained by
$\Sigma E_{\nu,i} n_{\nu,i} / \Sigma n_{\nu,i}$.
In Figure \ref{fig:numean_plot} we plot the averaged neutrino
energy of the three models for both $\nu_e$ and $\nu_{\tau}$
as a function of time. 

The $\nu_{{\tau}}$ has always a higher mean energy than the
$\nu_e$. The typical neutrino energy is $\sim$ 0.9 MeV
in the quiescent time, and increases to its peak $\sim$ 1.1 MeV
when the star is the most compact. The maximum mean energy
of neutrinos decreases when the He core mass increases
during the first peak. This is because when the He core
is more massive, the corresponding central density
becomes lower when the explosive O-burning is triggered.
In the $\rho$-$T$ diagram, the trajectory of the core
is closer to the pair-creation instability zone.


\subsection{Pre-collapse Neutrino Signal}

In this part we further examine the neutrino production of PPISN
before its collapse.
Unlike the pulsation, when the star finally runs out of $^{16}$O 
for its explosive burning, the core is sufficiently massive
that it promptly collapses. In this phase, although it 
can reach a higher central density and temperature, which is
favorable for neutrino emission, the respectively shorter
timescale also limits the number of neutrinos emitted. To demonstrate
the similarity of the pre-collapse in different models,
we consider the two contrasting models, namely 
Models He40A and He62A, to examine how the neutrino
number flux and the energy distribution vary with time.

\begin{figure*}
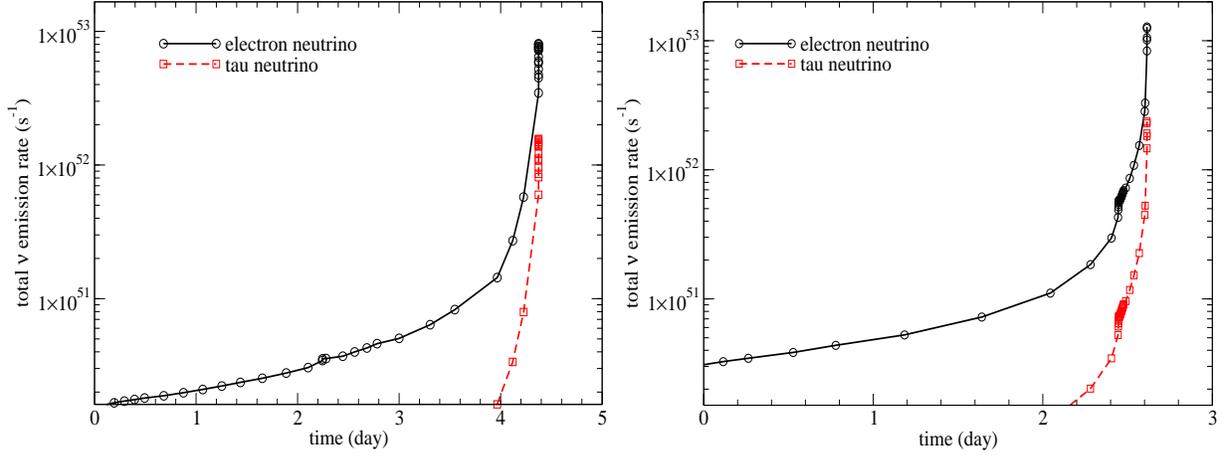

\centering
\includegraphics*[width=8cm,height=6cm]{fig6a.eps}
\includegraphics*[width=8cm,height=6cm]{fig6b.eps}
\caption{
(left panel) Neutrino number emission rate against time for
$\nu_e$ and $\nu_{\tau}$ before the onset of collapse of Model He40A.
(right panel) Same as the left panel, but for Model He62A.
In both panels, time 0 is shifted such that 
the relevant time range can be shown until the simulation ends.}
\label{fig:nunum_final}
\end{figure*}

In Figure \ref{fig:nunum_final} we plot the neutrino 
number emission rate against time for both $\nu_e$ and $\nu_{\tau}$
during the pre-collapse phase of the two models. 
The neutrino number emission becomes significant 
only at 0.001 -- 0.002 year ($< 1$ day) before 
the collapse. The two types of neutrinos can have their
number emission rates increased by 2 -- 3 orders
of magnitude, until their peaks of $\sim$ $10^{53}$ erg s$^{-1}$,
when the simulations stop. We do not evolve further
because beyond that,
nuclear physics and neutrino transport become important but
these physics components are not implemented in the stellar
evolution code, when the density exceeds $\sim 10^{11}$ g cm$^{-3}$.

\begin{figure*}
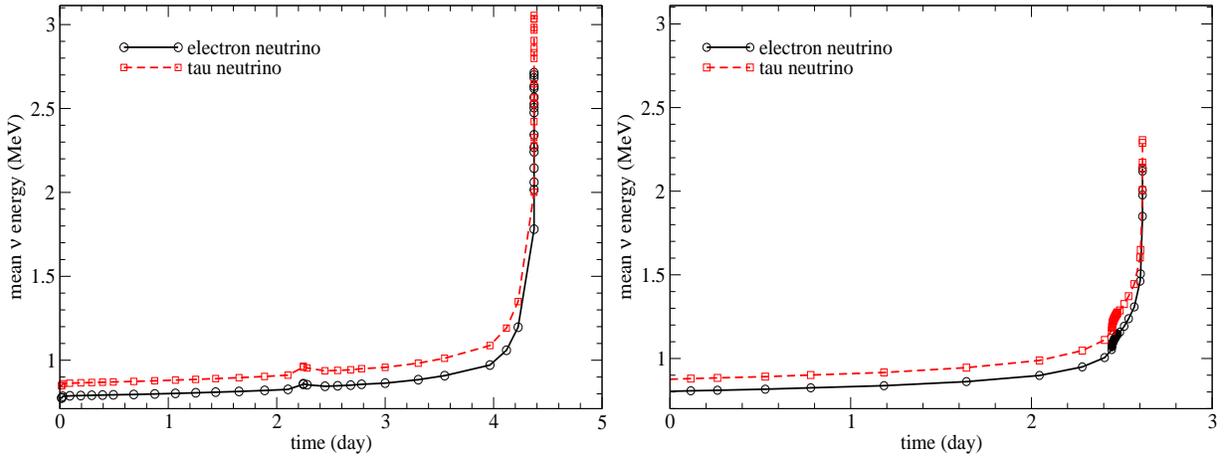

\centering
\includegraphics*[width=8cm,height=6cm]{fig7a.eps}
\includegraphics*[width=8cm,height=6cm]{fig7b.eps}
\caption{
(left panel) Mean neutrino energy against time for
$\nu_e$ and $\nu_{\tau}$ before the onset of collapse of Model He40A.
(right panel) Same as the left panel, but for Model He62A.
The time convention follows Figure \ref{fig:numean_final}.}
\label{fig:numean_final}
\end{figure*}

In Figure \ref{fig:numean_final} we plot the mean 
neutrino energy against time for the two models. 
Unlike the mean energy in the pulses, the 
mean energy for both types of neutrinos can be higher 
as a result of higher central temperature ($\sim 10^{10}$ K)
before collapse. This shifts the thermal spectra 
towards a higher energy, where at the peak the 
neutrinos can have an average energy of $\sim 3$ MeV.
No qualitative difference can be found 
between the two contrasting models.

\begin{figure*}
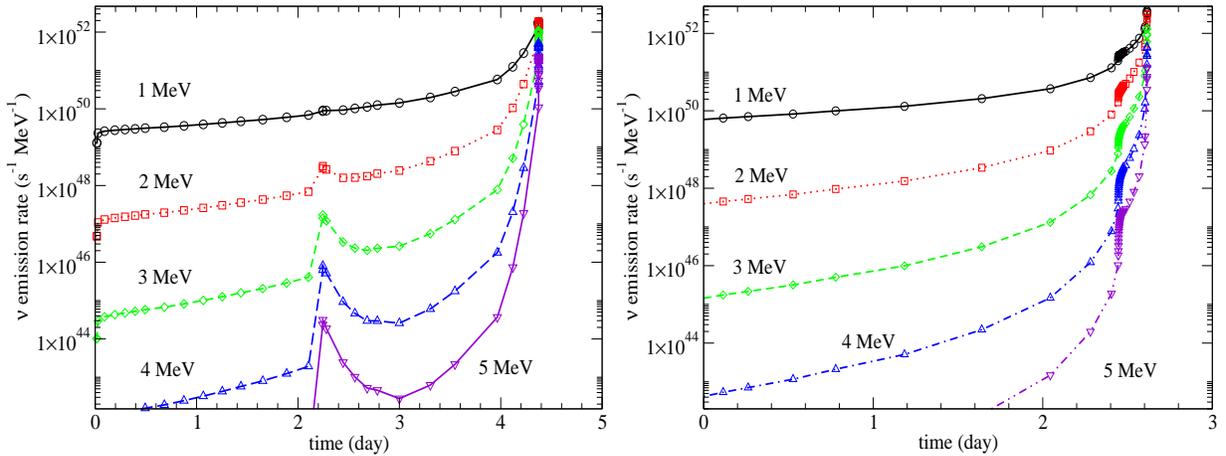

\centering
\includegraphics*[width=8cm,height=6cm]{fig8a.eps}
\includegraphics*[width=8cm,height=6cm]{fig8b.eps}
\caption{
(left panel) Neutrino spectral time evolution for
$\nu_e$ from 1 to 5 MeV before collapse in Model He40A.
(right panel) Same as the left panel, but for Model He62A.
The time convention follows Figure \ref{fig:numean_final}.}
\label{fig:nuspec_final}
\end{figure*}

In Figure \ref{fig:nuspec_final} we plot the spectral evolution
of the two models for neutrinos with an energy from 
1.0 MeV to 5.0 MeV. In the two models, a thermal-like
distribution can be observed. The high energy neutrino 
(5 MeV) can be comparable with the low energy neutrino
(1 MeV) only at the moment very close to the onset of collapse.

\section{Discussion}
\label{sec:discussion}

\subsection{Predicted neutrino signals}

\subsubsection{Neutrino Energy Distribution}

\begin{figure}
\centering
\includegraphics*[width=8cm,height=6cm]{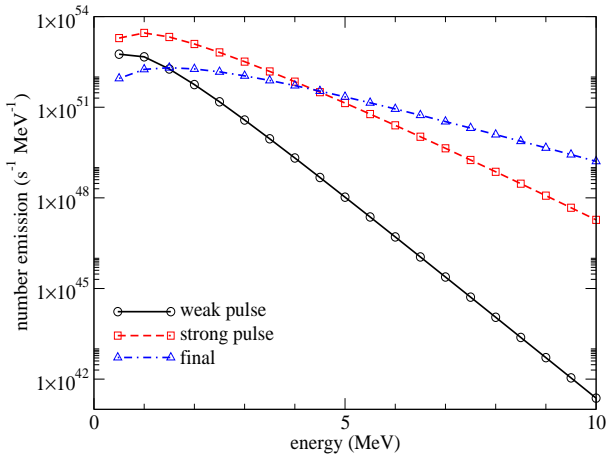}
\caption{Neutrino spectral snapshots of Model He40A
for three moments, at the neutrino emission peaks 
during the second pulse (weak pulse), the sixth pulse (strong pulse) 
and near the end of the simulation.
}
\label{fig:nuspectralsnaphot_plot}
\end{figure}

In this section we examine the expected neutrino 
signals by the terrestrial neutrino detectors. We examine how 
the neutrino energy distributions look like in all the three
cases. They include the neutrinos produced during 
the weak pulse, the strong pulse and 
in the pre-collapse phase. We want to examine if the energy distribution 
provides important indications that the neutrinos
detected comes from the PPISN, instead of other 
possible background. 

To illustrate the difference, we consider Model He40A
and take three spectral snapshots at three moments, 
when the star has a maximum neutrino emission 
(1) in the second pulse (weak pulse), (2) at the sixth 
pulse (strong pulse) and (3) near the end of simulation
(pre-collapse). They have neutrino number emission
rates at $9.16 \times 10^{52}$, $6.68 \times 10^{53}$ and
$8.08 \times 10^{52}$ s$^{-1}$ respectively. 
In Figure \ref{fig:nuspectralsnaphot_plot}
we plot the spectra of these three moments. 
In the weak pulse, where the 
star is not globally heated by the explosive O-burning,
the neutrino distribution is monotonically decreasing.
In the strong pulse, the energy spectrum shows the form $\sim \exp(-E_{\nu}/k_{\rm B}T)$.
There is a peak emission around 1 MeV
and then the emission rate quickly drops. There is a 
five-order-of-magnitude difference for neutrinos numbers between energy
of 1 MeV and  
of 5 MeV. At last, in the pre-collapse moment, 
although it has in total a lower neutrino emission,
the neutrino distribution extends to a higher energy.
The peak shifts to $\sim 2$ MeV, with the 5 MeV
neutrino being comparable with the lower energy
neutrinos. 

From this comparison it shows that, despite that the 
neutrino spectra are collections of all the 
fluid elements in the star, which have
a wide range of density and temperature,  
the overall spectra are still comparable to the Boltzmann distribution.
Furthermore, the low energy neutrinos
carry most of the thermal energy in the 
pulses, while neutrinos in a wider energy range
can be found in the pre-collapse scenario. 

In the above analysis we have assumed the neutrino 
directly reaches the Earth without any interaction.
In fact, the neutrino oscillation and the 
mass hierarchy of neutrino can play a role in the 
final neutrino count. The neutrino oscillation 
and resonances with leptons by Mikheyev-
Smirnov-Wolfenstein (MSW) effects may
further alter the original neutrino sources. 
The mass hierarchy changes the rate of oscillation
by its extra interaction term in the flavor eigenstate 
oscillation Hamiltonian.
However, as shown in \cite{Wright2017a},
the differences between the normal and inverted mass 
hierarchies are subtle. Given the 
uncertainties to the other parts of input physics,
we expect that the difference among different
mass models may be too small to be observed.  

To estimate the astrophysical origin, we assume 
the star to be at 1 kpc from the Earth. This stands for a
surface area about $1.20 \times 10^{44}$ m$^2$ for the 
neutrino flux.
We remind that in fact there exists massive stars near our
neighbourhood. In Table \ref{table:massive_stars} we tabulate 
some of the nearby stars which has a mass above 80 $M_{\odot}$
and has a distance around 1 kpc. These stars can be the 
candidates for the future pulsation events when their
He core mass grows to the mass range necessary for pair-instability.

\subsubsection{Neutrino Number Counts}

We have presented in Section \ref{sec:nu_signal} the detailed neutrino 
emission profiles and the history of the representative PPISN models.
Here we estimate the possible detection by 
terrestrial neutrino detectors. 
To estimate the detection counts, we use the following estimation.
We assume the detection relies on the weak interaction
$p(\bar{\nu}_{e},e^+)n$, where the positron is quickly
annihilated by surrounding electrons. 
The cross-section is given by
\begin{eqnarray}
\sigma = \frac{G_F^2 \epsilon_{\nu}^2}{(\hbar c)^4 \pi} (C_V^2 + 3 C_A^2) \left( 1 - \frac{Q}{\epsilon_{\nu}} \right) \times \\
\sqrt{1 - 2 \frac{Q}{\epsilon_{\nu}} + \frac{Q^2 - m_e^2}{\epsilon_{\nu}} \Theta(\epsilon_{\nu} - Q)},
\end{eqnarray}
where $C_V$ and $C_A$ are the vector and axial-vector
coupling constants, $G_F$ is the Fermi weak coupling constant,
and $\epsilon_{\nu}$ is the neutrino energy. $Q = 1.3$ MeV is 
the mass-energy difference between $p$ and $n$ and 
$m_e = 511$ keV is the electron mass. 
The step function arises naturally from the mass 
difference between $n$ and $p$ such that the interaction
occurs only when the neutrino is sufficiently energetic.
We assume that the canonical neutrino detector contains water
of a mass 10 kton. This represents an amount of 
$\sim$ $6.69 \times 10^{32}$ hydrogen atoms.

In Table \ref{table:nurate_detector}, we tabulate the optimistic detection numbers 
for different current and proposed neutrino detectors. Rates 
below 1 count per minute is neglected. The neutrino source 
is assumed to be at 1 kpc. 
We assume a uniform energy bin of 0.5 MeV
from 0.5 to 20 MeV. Due to the Heaviside function the neutrino
below $\sim 1.5$ MeV is cut off by default. We further assume a perfect
detection rate for the neutrino detector. We do that because
the actual detection accuracy depends on the energy threshold,
the detection acceptance rate and the energy reconstruction algorithm
of individual neutrino detectors. 
However, not all data is openly available. 
The energy threshold, in particular, is detector-dependent.
For example, LENA \citep{Wurm2015} is proposed to have a threshold energy
as low as 2 MeV. On the other hand, the threshold energy for ICECUBE 
can be as high as 200 TeV \citep{Aartsen2016}. The incoming 
neutrinos from PPISN will be shielded by noise in ICECUBE, but they can be detected by LENA. 
As a first approximation we assume the detector has a perfect detection rate.
The current neutrino detectors such as KamLAND, SNO+ and Borexino
are on the lower side of detection counts that except Model He40A,
neutrinos emitted from more massive star models 
in general cannot detect any significant number of neutrinos.
Future neutrino detectors such as JUNO and LENA can detect more
neutrinos in the order of $O(10)$. 
Super-Kamiokande and Hyper-Kamiokande can predict the 
highest amount of neutrinos from $\sim 10$ to $\sim 100$.

\begin{table*}
\begin{center}
\caption{The nearby massive stars which have a distance below 
10 kpc and a mass above 80 $M_{\odot}$. Mass
is in unit of $M_{\odot}$. Distance is in unit of kpc.}
\label{table:massive_stars}
\begin{tabular}{|c|c|c|c|}
\hline
Star & Mass & Distance & Reference \\ \hline
Cygnus OB2-12 	& 110			& 1.6 & \cite{Oskinova2017,Camarillo2018} \\ \hline
HD 93129 A 		& 110			& 2.3 & \cite{Cohen2011} \\ \hline
$\eta$ Carinae A 	& $\sim 100$	& 2.3 & \cite{Walborn2012,Kashi2010} \\ \hline
Cygnus OB2 \#516 & 100			& 1.4 & \cite{Herrero2002} \\ \hline

\end{tabular}
\end{center}
\end{table*}

\begin{table*}

\begin{center}
\caption{The optimistic total neutrino number detection count to be received by
terrestrial neutrino detectors. Masses of the detector are in units of
kT.
The star model is assumed to be at 1 kpc from the 
Earth. We refer the readers to Paper I
for the detailed description of each pulse.}
\label{table:nurate_detector}
\begin{tabular}{|c|c|c|c|c|c|c|c|c|c|c|}
\hline
Model	 & Mass  	& He40A   & He45A   & He50A   & He55A   & He55A   & He60A   & He60A & He62A & He62A \\ \hline
Pulse	 &           & 1-6     & 1-4     & 1-2     &       2 &     3   & 1       & 2     & 1     & 2     \\ \hline
KamLAND						& 1.0 & 2 & $<1$ & $<1$ & $<1$ & $<1$ & $<1$	& $<1$ &$ <1$ & $<1$ \\
SNO+ 						& 0.78 & 1 & $<1$	& $<1$ & $<1$ & $<1$ & $<1$	& $<1$ & $<1$ & $<1$ \\
Borexino 					& 0.278 & $<1$ & $<1$ & $<1$ & $<1$ & $<1$ & $<1$	& $<1$ & $<1$ & $<1$ \\
JUNO 						& 20 & 36 & 8 & 4 & 2 & 2 & $<1$ & 1 & $<1$ & 1 \\
RENO-50						& 18  & 32 & 7	& 3 & 2 & 1 & $<1$	& 1 & $<1$ & 1 \\
LENA 						& 50 & 90 & 20 & 9 & 6 & 4 & 1	& 3 & 1 & 3 \\
Super-Kamiokande (with Gd) 	& 32.5 & 40 & 9	& 4  & 3  & 2  & $<1$	& 1 & $<1$ & 1 \\
Hyper-Kamiokande (with Gd) 	& 220 &  680 & 150	& 72 & 43 & 29 & 7	& 20 & 6 & 20 \\
DUNE					& 40 & 36 & 8 & 4 & 2 & 2 & $<1$ & 1 & $<1$ & 1 \\
 \hline

%

\end{tabular}
\end{center}
\end{table*}

\begin{figure}
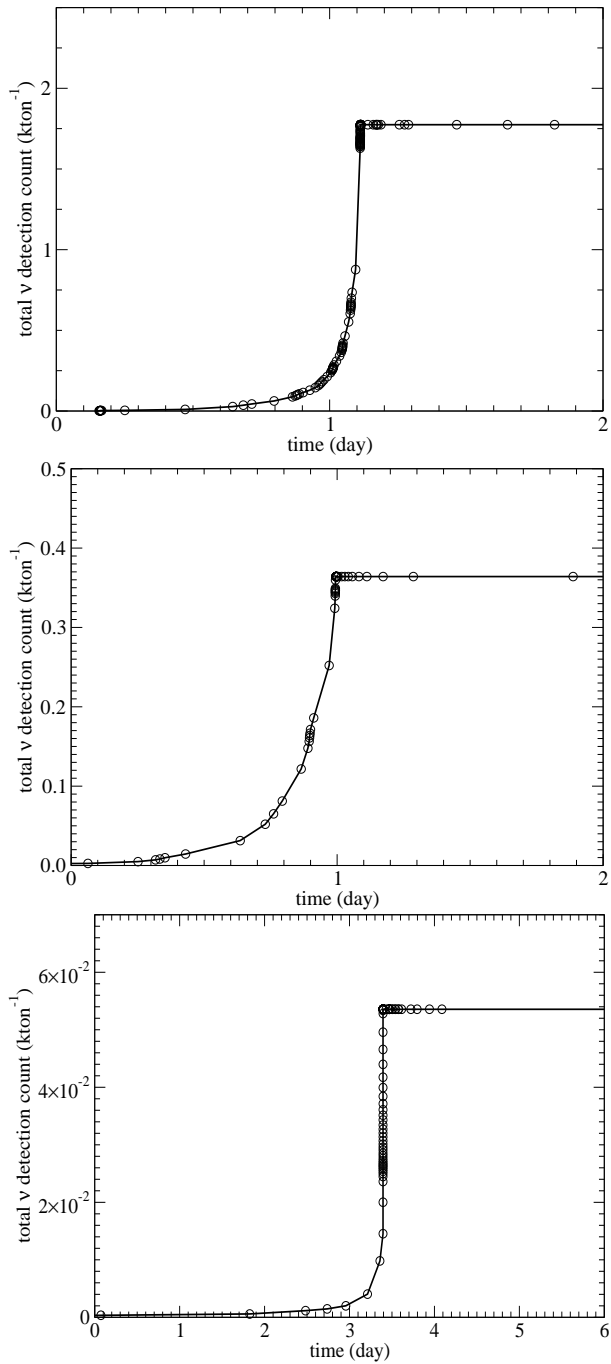

\begin{center}
\includegraphics*[width=8cm,height=6cm]{fig9a.eps}
\includegraphics*[width=8cm,height=6cm]{fig9b.eps}
\includegraphics*[width=8cm,height=6cm]{fig9c.eps}
\caption{
(top panel) Cumulative $\nu_e$ count against time for
$\nu_e$ and $\nu_{\tau}$ of the Model He40A. Notice that the 
unit here is kton$^{-1}$. 
The neutrino detector is assumed to have a mass of 1 kton 
and located at 1 kpc from the supernova. 
(middle panel) Same as top panel, but for Model He50A.
(bottom panel) Same as top panel, but at the peak of 
the second pulse for Model He62A.
Time zero is shifted so that the relevant time period
can be shown directly.}
\label{fig:nucount_plot}
\end{center}
\end{figure}

Based on the above methods, in Figure \ref{fig:nucount_plot},
we plot the cumulative $\nu_e$ count of each strong pulse for 
Models He40A, He50A and He62A per 1 kton of the detecting 
material for an astrophysical source at a distance of 1 kpc.
The cumulative sum is assumed to count across each 
pulse individually. Based on the number of strong pulses
experienced in the models, the cumulative counts differ
slightly. Most
neutrinos are detected within 0.002 year ($\approx$ day).
The following expansion of the supernova no longer
produces an observable amount of neutrino. To connect with the results in 
Table \ref{table:nurate_detector}, we need to multiply the results
in the figure by the mass of the neutrino detector and divide
the distance squared in unit of kpc.

For a lower mass He core (40 -- 55 $M_{\odot}$), there is only one 
strong pulse, as a result, the core tends to be more compact
when it stops contraction and starts its expansion. 
The typical density of the star is higher,
thus allowing more neutrinos to be generated. The
total number detected by the model neutrino detector, 
assumed to be 1 kpc away from the supernova and has
a detection mass of 1 kton, is higher. It has a 
typical value of $O(10^1)$ across the pulsation,
and the total number decreases with mass. 

For a higher mass He core (55 -- 62 $M_{\odot}$), there are two strong pulses.
The first pulse occurs very soon in the contraction phase
because of the abundant $^{16}$O in the core. 
Therefore, the corresponding density and temperature
of the star is lower. 
The typical neutrino count is lower, 
$\sim 10^6$ kton$^{-1}$ across the event. 
On the other hand, in the second pulse, 
because the core has much less $^{16}$O than the first pulse,
the core needs to reach a more compact state during
contraction, in order to make the outer core
where $^{16}$O is not yet burnt during the 
first pulse. There are more neutrinos
detected during the pulsation. Despite that, the total
neutrinos detected are still less than those from its lower mass counterpart.

\subsection{Comparison with Other Types of Supernovae}

The possibility of using neutrinos as a precursor to detect 
the emergence of a supernova has been proposed in the literature.
The early light curve can provide important information
about the outer structure of the star, which cannot be easily
detected \citep{Bersten2018}. It occurs very soon 
after the explosion, in the scale of shock crossing
time of the envelope. It requires coincidences
for orienting the telescope to the supernova hosting 
galaxy right at the moment where the explosion starts,
if no early warning signal is provided.  
On the contrary, when the associated neutrinos can be detected,
there is a time delay between the arrival of neutrino
and photons. The shock propagates at a sub-light speed
velocity towards the surface, compared with the neutrinos
traveling in the speed of light. The difference can 
be varying from a few seconds (for a Type Ia supernova),
to a few minutes (for a blue supergiant), and up to as much as a 
$\sim$ hours (for a red supergiant). See for example
\cite{Dessart2017,Owocki2019} for recent theoretical
predictions of shock breakout in massive stars
and \cite{Garnavich2016} for a recent observation
of the early time light curve demonstrating shock breakout
in a massive star explosion.
The optical 
evolution of these shock breakout events contains very useful
information about the pre-explosion structure of the 
star. Besides, the neutrinos detected contain
information directly from the stellar core.

In Table \ref{table:neutrino_compare} we compare
the neutrino luminosity, energy and detection
counts for different types of supernovae. 
$E_{\nu}$ is the average neutrino energy
per particle and $L_{\nu}$ is the total neutrino luminosity.

Type Ia supernova, as an explosion by thermal nuclear
runaway in a carbon-oxygen or oxygen-neon-magnesium 
white dwarf, can generate neutrino by 
both thermal processes and electron captures. 
Thermal processes include such as pair-neutrino 
in the thermalized core, in particular in regions
where complete burning proceeds (burning of matter
until nuclear statistical equilibrium is reached). Electron
captures occur mostly in the burnt matter in NSE
with a high density ($\sim 10^9$ g cm$^{-3}$).
In this density range, electrons become extremely degenerate
with a high Fermi energy, which may exceed the 
mass-energy difference between a neutron and a proton.
This favours the capture of electrons on the nuclei
and results in $\nu_e$ emission. Computation of 
electron capture in these supernova is important 
for a self-consistent computation. See for example
\cite{Seitenzahl2009} for the local electron capture rate for stars
undergoing thermonuclear explosions and 
\cite{Jones2016,Leung2018,Leung2019ECSNReview,Leung2019ECSN} 
for recent SN Ia 
simulations including electron captures.
Depending on the explosion mechanisms, the runaway
can propagate in the form of sub-sonic deflagration
or supersonic detonation. In both cases, the burnt ash
can reach the temperature $\sim 5 - 9 \times 10^9$ K,
where matter achieves the nuclear statistical equilibrium.
In \cite{Odrzywolek2011},
\cite{Leung2015a} and \cite{Wright2017a}, 
the neutrino productions 
are analyzed for the pure turbulent deflagration (PTD), 
turbulent deflagration model with deflagration-detonation
transition (DDT) and gravitationally confined detonation (GCD) models. 
The time evolution of neutrino is sensitive to the 
explosion mechanism, for example the one-peak structure 
for the PTD model versus the two-peak structure in DDT
and GCD model. In all models they have the lowest 
neutrino luminosity and possible counts in major
representative neutrino detectors. But they 
have an intermediate averaged neutrino energy. 

A PISN also shares a similar neutrino production mechanism
because of its thermonuclear origin. Different from a
Type Ia supernova, the much more massive hot core $2 - 4 ~M_{\odot}$,
can generate more $^{56}$Ni before it is completely
disrupted. In \cite{Wright2017b}, the 
neutrino emission signal is also analyzed. 

Core-collapse supernova has a completely different
neutrino production mechanism by electron capture and
neutron star cooling processes, such as the URCA process. 
Prior to its collapse, the deleptonization
via $e^- + ^{A}_{Z}X \rightarrow ^{A}_{Z-1}X' + \nu_e$
and thermal neutrinos contribute to neutrino cooling. 
The thermally excited core is also about $10^9 - 10^{10}$ K
when the core reaches $10^{10}$ g cm$^{-3}$. 
In \cite{Yoshida2016b} the neutrino signals from 
12 -- 20 $M_{\odot}$ stars are studied. The neutrino generation
is in general monotonic increasing in time before its collapse. 
Massive star has a lower neutrino luminosity but 
still a significant detection count. It is because the
pre-collapse phase includes also the hydrostatic Si-burning,
which can take place $\sim 1$ day when the core reaches
$> 10^9$ K before collapse.

We remark that even though the Type Ia supernova explodes in 
a similar manner comparable to the pulsation mechanism 
in PPISN and also PISN, 
it has a much lower detection count for three reasons. 

First, the mass inside the star which can efficiently generate neutrino,
in particular the pair-neutrino, is much lower than the latter
two cases. The maximum mass it can incinerate is the 
Chandrasekhar mass ($\sim 1.4 ~M_{\odot}$) or about 
1 $M_{\odot}$ for the sub-Chandrasekhar mass case. 
On the other hand, in a PISN or a PPISN, the amount of
mass capable of incinerating $^{16}$O and reaches
above $10^9$ K can range from a few to $\sim$ 30 $M_{\odot}$.

Second, the timescale for the star to emit 
neutrino is much longer in a PPISN and a PISN, compared to
a Type Ia supernova. In a Type Ia supernova, 
from the incineration to the expansion,
the time duration where the matter reaches 
the temperature above $10^9$ K is less than 
1 -- 2 s, which is the typical time for the 
deflagration and detonation wave to swept across
the star and to disrupt the star. On the other hand, due to a longer
dynamical timescale ($\geqslant$ 100 -- 1000 s), 
the total number of neutrinos emitted 
by PPISN and PISN can be much higher. 

Third, the typical density in Type Ia supernovae is 
much higher in the Chandrasekhar mass scenario. 
The central density is about $10^9$ g cm$^{-3}$
(although variation exists as indicated from 
different Type Ia supernova observations \citep{Leung2018, Nomoto2017}
and from the progenitor \citep{Nomoto2018}).
The strong degeneracy limits the emission rate.
Notice that the thermal neutrinos can also be emitted
strongly during the nuclear runaway phase in 
the electron capture supernova \citep{Nomoto1988, Doherty2015, 
Leung2017MMSAI, Leung2019ECSNReview} (Also applies for ONeMg core).
Before the star collapses into a neutron star, the 
O-Ne deflagration also allows the matter to reach 
$\sim 10^9$ K \citep{Leung2019ECSN}. Furthermore, the 
pre-runaway electron captures by $^{20}$Ne and 
$^{24}$Mg provide another channel for producing
neutrinos besides thermal neutrinos \citep{Nomoto2017HBSN-ECSN, Toshio2019, Zha2019}. 

\begin{table}
\begin{center}
\caption{The typical neutrino properties from different
types of supernovae. Neutrino luminosity at peak $L_{{\rm peak}}$ is in unit
of erg s$^{-1}$ and neutrino energy $E_{{\rm \nu}}$ is in unit of 
MeV. Neutrino Count $N_{\nu(i)}$ is the number of neutrino expected
to be detected by Super-Kamiokande $(i) = (S)$ and Hyper
Kamiokande $(i) = (H)$ when the explosion occurs at 10 kpc away
from the Earth. For Type II supernova, we only choose the neutrino
luminosity and energy before its collapse for a better comparison 
with PPISN and PISN where thermal neutrino is the main component.
Massive star includes the neutrino emission before the 
onset of its Fe-core collapse.}

\label{table:neutrino_compare}
\begin{tabular}{|c|c|c|c|c|}
\hline
Supernova & $L_{{\rm peak}}$ & $E_{\nu}$ & $N_{\nu(S)}$ & $N_{\nu(H)}$ \\ \hline
Type Ia (PTD) \footnote{\cite{Odrzywolek2004}} & $10^{49}$ & 3.8 & 0.063 & 0.106 \\
Type Ia (DDT) \footnote{\cite{Odrzywolek2004}} & $10^{49}$ & 3.5 & 0.013 & 0.220 \\
Type Ia (GCD) \footnote{\cite{Wright2017a}} & $10^{47}$ & 0.5/3 & 0.0024 & 0.0267 \\
ONeMg core \footnote{\cite{Kato2015}} & $10^{46}$ & 1 - 2 & $< 1$ & $< 1$ \\
Massive star ($M = 15 M_{\odot}$) \footnote{\cite{Yoshida2016}} & $10^{47}$ & 2.0 & 15 & 250 \\
PPISN ($M_{{\rm He}} = 40 M_{\odot}$) & $10^{47}$ & 1.5 & 0.403 & 6.80 \\
PPISN ($M_{{\rm He}} = 62 M_{\odot}$) & $10^{47}$ & 1.0 & 0.0102 & 0.203 \\
PISN ($M = 250 M_{\odot}$) \footnote{\cite{Wright2017b}} & $10^{50}$ & 2 & 6.98 & 52.23 \\ \hline

\end{tabular}
\end{center}
\end{table}

\subsection{Conclusion}

In this article we extended our previous study of pulsation
pair instability supernovae to examine the associated neutrino signals. In \cite{Leung2018} we have 
performed one-dimensional stellar evolutionary simulations of this class
of supernovae using the one-dimensional stellar evolution 
code MESA version 8118. We followed the evolution of the He core
since the main-sequence phase until the collapse of the star. 
Meanwhile, we recorded the thermodynamics trajectories of
the star for analyzing its neutrino emission done
in this work.

We use the neutrino subroutine $sneut5$\footnote{open source subroutine available
on http://cococubed.asu.edu/code\_pages/nuloss.shtml. The subroutine
summarizes the parametrized neutrino emission rates in works including
\cite{Itoh1996}.} for calculating the
detailed neutrino emission of the He core models 
of mass 40 -- 64 $M_{\odot}$. We follow their neutrino
emission history from the onset of pulsation
until its collapse. 
We further 
extract its spectra by the semi-analytic formulae
of pair- and plasmon-neutrinos. 
We analysed the possible neutrino observables for He cores
from 40 to 64 $M_{\odot}$. They correspond to the main-sequence
stars of masses $\sim$ 80 -- 140 $M_{\odot}$ (but with metallicity
dependence). We find that neutrinos are mostly produced
by the pair-neutrino channel ($e^- + e^+ \rightarrow \nu_e + \nu_{\bar{e}}$).
Most of these neutrinos are emitted within the one hour 
during its contraction prior to its pulsational mass loss. 
The lower mass star tends to emit more
neutrinos and has higher detection counts because it is more compact. 
Due to its thermal nature, the neutrinos have an averaged energy
about a few MeV. At last, using the pair-neutrino as
an example, we confirm that the current analytic approximation 
formula of neutrino production \citep{Itoh1994} can 
well match the more updated neutrino luminosity 
table given in \cite{Odrzywolek2007}.

This work shows that the repeated pulsations 
of PPISN allow the star to reach the hot and compact
state more frequently than its more massive relative 
(pair instability supernova) and less massive 
relative (core collapse supernova). This provides
more opportunities in predicting its collapse 
by detecting its neutrinos. 
Future detection of 
these neutrinos may serve as an early warning 
signals for the optical telescopes to detect the electromagnetic wave
signals coming from the early shock breakout.
Those neutrinos contain precious information about the 
pre-explosion stellar structure.

\section{Acknowledgment}

This work has been supported by the World Premier International
Research Center Initiative (WPI Initiative), MEXT, Japan, and JSPS
KAKENHI Grant Numbers JP17K05382 
and 26104007 (Kakenhi).
S.B. work on PPISN is supported by the Russian Science
Foundation Grant 19-12-00229.
We thank F. X. Timmes for
his open-source micro-phsyics algorithm
including the Helmholtz equation of state 
subroutine and the neutrino subroutine
$sneut5$. We also thank A. Odrzywolek for 
supplying the open-source pair-neutrino 
table for cross-checking with other approximation
formula. We also thank Professor 
Mark Vagins for the informative introduction
on the neutrino detection techniques and 
guidance in the Super-Kamiokande and KamLAND detection site.

\appendix

\section{Use of Analytic Approximation for Neutrino Luminosity}
\label{sec:analytic}

In the main text we have studied the neutrino emission 
based on the implicit subroutine included in MESA for the 
neutrino light curve \footnote{http://cococubed.asu.edu/code$\_$pages/nuloss.shtml}
and some analytic approximations 
for the neutrino spectra \citep{Odrzywolek2007}.
The subroutine summarized the analytic approximations
presented in \cite{Itoh1994}, with the detailed 
calculation described in \cite{Itoh1989}.
The subroutine $sneut4$ and $sneut5$ correspond to
the same input physics but for the single and double 
precision. The subroutine has been widely applied to many 
applications in stellar astrophysics. However, with the 
more detailed calculations in some of the neutrino
processes \cite[e.g.][]{Odrzywolek2007,Misiaszek2006}, 
it is unclear whether this approximation remains fully accurate. 
To check its accuracy, we compare the neutrino luminosity 
from pair-production. This process is the most important
neutrino production channel for massive stars due to its
low density-high temperature core. To compare with, 
we use the neutrino table\footnote{http://th.if.uj.edu.pl/$\sim$odrzywolek/psns/index.html}
and the analytic formula given in \cite{Blinnikov1989}. 

\begin{figure}
\centering
\includegraphics*[width=10cm,height=8cm]{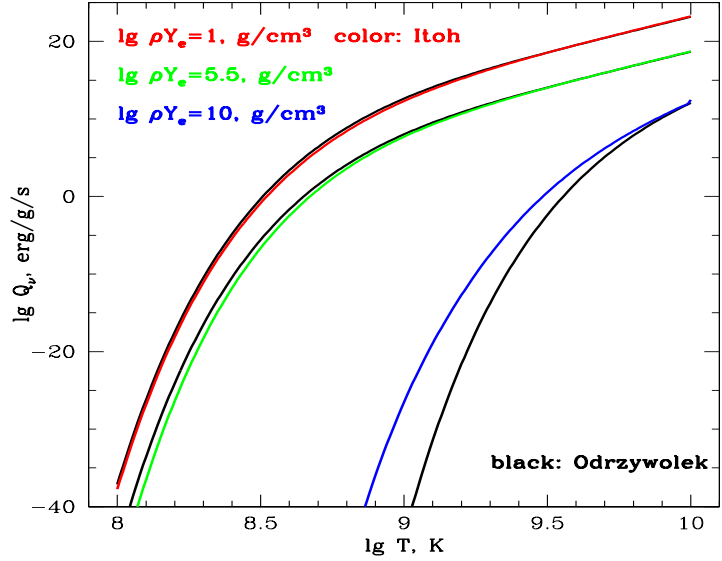}
\caption{Comparison between the analytic formula
for the pair-neutrino production used in the MESA code and the 
table based on \cite{Odrzywolek2007}. The black line
corresponds to the numerical data from the table
and the lines of other colours correspond to those
from the analytic formula.}
\label{fig:odryitoh}
\end{figure}

In Figure \ref{fig:odryitoh} we compare the pair-neutrino
luminosity at different densities from $1 - 10^{10}$ g cm$^{-3}$
and different temperature $10^8 - 10^{10}$ K. At a low density, 
the two curves overlap with each other, showing that
at low density-high temperature regime, the analytic formula
is a very good approximation compared to the exact values
presented in table form. This is important because this is
a typical temperature 
and density similar to that during the pair-creation instabilities in
most stellar models. This guarantees
the accuracy of neutrino energy loss in the pre-supernova 
evolution. 

In an intermediate density ($\sim$ $10^{5.5}$ g cm$^{-3}$, 
the two curves still overlap well except at 
low temperature around $10^8$ K, where the 
discrepancy is within one order of magnitude.
Above $10^9$ K, the formula agrees very well
with the table. 
We remark that at that density range, the 
pair neutrino is less important. 

At high density ($\sim$ $10^{10}$ g cm$^{-3}$), the discrepancy
becomes much larger at low temperature. The discrepancy
is less severe at temperature $10^{9.5}$ K but below
that, the error grows when temperature drops. 
The discrepancy can be as large as ten orders
of magnitude. 
Again, the large discrepancy does not affect
the total neutrino calculation because at such 
high density, the photo-neutrino and electron
bremmstrahlung are the major channels for the neutrino
production. 

From the three regimes, it suffices to conclude that
for the current neutrino calculation, the analytic 
approximation of the pair-neutrino mechanism can very well
describe the neutron luminosity.

At last we apply this comparison to a specific stellar profile
obtained from our calculations. We use the Model He60A
as an example. We input the temperature, density and
composition obtained from the profile, and then 
compare the corresponding pair-neutrino luminosity 
at different positions in the star. The profile is taken from
the model when the star obtained its highest central
density during the first pulse. In the core (within 
zone 400), due to the high central density, 
the electron matter becomes degenerate, which suppresses 
the pair-neutrino. Despite that, the high temperature in 
the core provides the condition where the analytic formula 
agrees very well with the table values. Outside the 
core, when the pair-neutrino becomes important, 
the two methods still agree well with each other. 
This shows that in the typical stellar calculation, 
the analytic approximations can still very well reproduce
the neutrino luminosity calculated from more accurate
ones by direct table interpolation.

\begin{figure}
\centering
\includegraphics*[width=10cm,height=8cm]{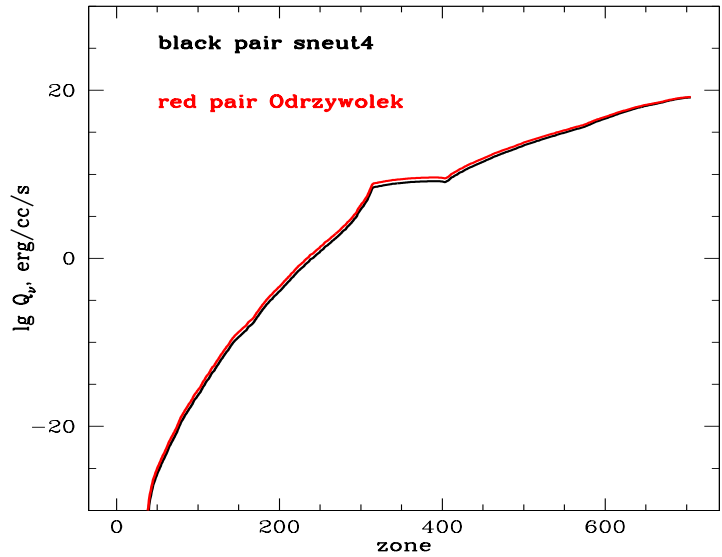}
\caption{The comparison between the analytic formula (black line)
for pair-neutrino used in the MESA code and the 
table (red line) based on \cite{Odrzywolek2007} for the 
PPISN model He60A. The stellar profile is 
obtained at the most compact state
in the first pulse.}
\label{fig:nuv}
\end{figure}

\bibliographystyle{apj}
\pagestyle{plain}
\bibliography{biblio}

\end{document}